\documentclass[useAMS,usenatbib,usegraphicx]{mn2e}
\usepackage{txfonts}
\title[Optical Follow-up of New SMC Wing Be/X-ray Binaries]{Optical Follow-up of New SMC Wing Be/X-ray Binaries}
\author[M.P.E. Schurch et al.]{M.P.E. Schurch$^{1}$\thanks{E-mail:
mpes@astro.soton.ac.uk},  M.J. Coe$^{1}$, K.E. McGowan$^{1}$, V.A. McBride$^{1}$, D.A. Buckley$^{2}$, 
\newauthor J.L. Galache$^{3}$, and R.H.D. Corbet$^{4}$\\
$^{1}$School of Physics and Astronomy, Southampton University, Highfield, Southampton, SO17 1BJ, UK.\\
$^{2}$South African Astronomical Observatory, P.O. Box 9, Observatory, 7935, South Africa.\\
$^{3}$Harvard-Smithsonian Center for Astrophysics, 60 Garden Street, Cambridge, MA 02138, USA.\\
$^{4}$Universities Space Research Association, X-ray Astrophysics Laboratory, Mail Code 662, NASA Goddard Space Flight Center, Greenbelt, MD 20771, USA.}
\begin{document}

\date{1 Aug 2007}

\pagerange{\pageref{firstpage}--\pageref{lastpage}} \pubyear{2007}

\maketitle

\label{firstpage}

\begin{abstract}
We investigate the optical counterparts of recently discovered Be/X-ray binaries in the Small Magellanic Cloud.  In total four sources, SXP101, SXP700, SXP348 and SXP65.8 were detected during the {\it Chandra} Survey of the Wing of the SMC.  SXP700 and SXP65.8 were previously unknown.  Many optical ground based telescopes have been utilised in the optical follow-up, providing coverage in both the red and blue bands.  This has led to the classification of all of the counterparts as Be stars and confirms that three lie within the Galactic spectral distribution of known Be/X-ray binaries.  SXP101 lies outside this distribution becoming the latest spectral type known.  Monitoring of the H$\alpha$ emission line suggests that all the sources bar SXP700 have highly variable circumstellar disks, possibly a result of their comparatively short orbital periods.  Phase resolved X-ray spectroscopy has also been performed on SXP65.8, revealing that the emission is indeed harder during the passage of the X-ray beam through the line of sight.
\end{abstract}


\section{Introduction}
Over the past 10 years the number of known high-mass X-ray binaries (HMXBs) located in the Small Magellanic Cloud (SMC) has been steadily increasing.  There are now approximately 50 such systems known in the SMC \citep{coe05}. Until recently the majority of X-ray observations by {\it Chandra}, {\it XMM-Newton} and {\it RXTE} have concentrated on the Bar, since the large fraction of HMXBs are located in this region.  \citet{coe05} analysed the locations of the known X-ray pulsars and believed there to be a relationship between the HI intensity distribution and the distribution of the pulsars. This prompted a large survey of the wing of the SMC using {\it Chandra}. A total of 20 individual pointings were made of approximately 10\,ks each.  The survey was successful in finding two new pulsars (SXP65.8 and SXP700) and detecting two previously known pulsars (SXP101 and SXP348).  Due to {\it Chandra's} high spatial resolution the counterparts to all four sources were identified (SXP348's being previously known).  Table~\ref{tasources} shows the positions and V band magnitudes of the counterparts to the detected pulsars. A more detailed description of the survey can be found in \citet{mcg07}.  In order to evaluate where these new pulsars lie in relation to the Galactic spectral distribution of Be/X-ray binaries we need to classify the counterparts. This paper reports the follow-up optical observations of these systems.

\begin{table*}
 \centering
 \begin{minipage}{140mm}
  \caption{New HMXBs in the Wing \label{tasources}}
  \begin{tabular}{@{}lllrr@{}}
  \hline
   SXP ID & X-ray and Optical Names & Coordinates & V Mag & Classification\\
 \hline
 SXP101 & RX J0057.3-7325, MACS J0057-734 10  & 00:57:27.08 -73:25:19.5 & 14.9$^{1}$ & B3-B5 Ib-II\\
 SXP700 & CXOU J010206.6-714115, [MA93] 1301  & 01:02:06.69 -71:41:15.8 & 14.6$^{2}$ & B0-B0.5 III-V\\
 SXP348 & SAX J0103.2-7209, [MA93] 1367       & 01:03:13.94 -72:09:14.4 & 14.8$^{3}$ & B0-B0.5 III-V\\
 SXP65.8 & CXOU J010712.6-723533, [MA93] 1619 & 01:07:12.63 -72:35:33.8 & 15.0$^{1}$ & B1-B1.5 II-III\\
 \hline
 \end{tabular}\\
$^{1}$\citet{mcg07}, $^{2}$\citet{mas02}, $^{3}$\citet{hug94}
\end{minipage}
\end{table*}

\section{Optical Data}
Since the {\it Chandra} observations were completed in March 2006 a number of telescopes have been used for the optical follow-up. Table 2 presents a list of the data collected for each of the sources.  The telescope configurations and reduction processes used are as follows: 
\begin{itemize}
\item AAT - 3.9m telescope, Anglo-Australian Observatory, Australia. The AAOmega optical spectrograph was used.  It is fed by the 2 Degree Field (2dF) robotic fibre positioner covering a 2 degree field at prime focus with 392 fibres. Each observation consists of three 1800\,s exposures made on 31 August 2006. The data reduction was performed using the latest 2dfdr package version 3.46. All traces were extracted using a tram line optimisation.  The spectra were then corrected for the detector response and then the red and blue arms were scaled and stitched together by forcing them to meet at 5900\AA\ allowing the complete range to be viewed, this was all performed using tasks within the AAT package.  The red arm has a dispersion of 1.5\AA/pixel and the blue of 1.03\AA/pixel.  The total wavelength range covered is approximately 3700\AA-8800\AA.  There are several bad columns in the AAT CCD.  These columns have been removed and appear as spaces in the spectra.

\item SALT - 11m Southern African Large Telescope, SAAO, South Africa. The Robert Stobie Spectrograph was used in conjunction with two gratings in long slit mode.
 \begin{itemize}
  \item Blue : 2300\,l/mm grating at an angle of 30.125$^{\circ}$.  This gives a dispersion of 0.34\AA/pixel and a wavelength range of approximately 3800\AA-4880\AA.
  \item Red : 1800\,l/mm grating at an angle of 36.5$^{\circ}$.  This gives a dispersion of 0.40\AA/pixel and a wavelength range of approximately 5930\AA-7210\AA.
 \end{itemize}
 
The SALT detectors consist of three chips aligned along the dispersion axis each separated by a small gap (evident in the spectra).  Full data reduction and mosaicking of the three chips were performed using the SALT IRAF pipelines and the standard IRAF packages in version 2.12.1 provided by the NOAO.  No flux calibration has been made.

\item ESO - 3.6m telescope, La Silla, Chile. The EFOSC2 faint object spectrograph was used combined with a grism with 600\,l/mm. The data reduction was performed using standard IRAF packages.  The spectra have a dispersion of 1.99\AA/pixel and a wavelength range of approximately 3080\AA-5100\AA, although the useful range is only from 3700\AA.  

\item SAAO - 1.9m telescope, SAAO, South Africa. The unit spectrograph was used combined with a 1200~l/mm grating and the SITe detector.  Data reduction was performed using standard IRAF packages. The dispersion of the spectra is fairly consistent over the years since the same set up has been used, approximately 0.43\AA/pixel. The wavelength range has varied slightly due to small changes in the grating angle, the wavelength range covered in all spectra is approximately 6230\AA-6970\AA.  A few cosmic rays have been removed by hand.

All the spectra except for those from ESO (due to their lower resolution) have been smoothed with a boxcar average of 5.  An arbitrary offset has been applied to some of the spectra to allow visual comparisons of the spectral lines.  
\end{itemize}

\begin{figure}
 \includegraphics[width=60mm, angle=90]{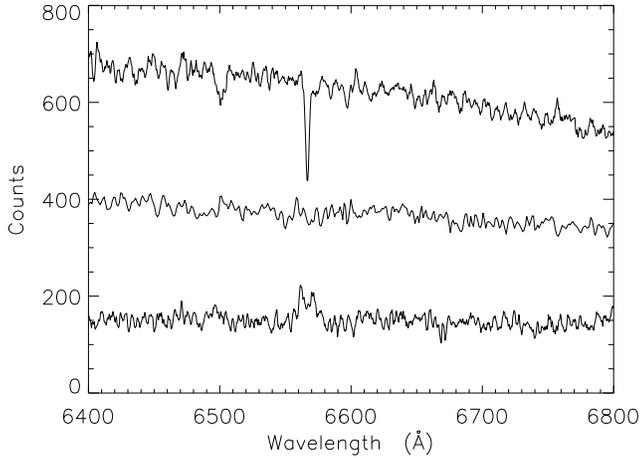}
  \caption{H$\alpha$ spectra of SXP101. Spectra dates top to bottom: 07 Sep 2004, 31 Aug 2006, 10 Nov 2006.\label{fi101ha}}
\end{figure}

\begin{table*}
 \centering
 \begin{minipage}{140mm}
  \caption{Observations \label{taobs}}
  \begin{tabular}{@{}llrrrr@{}}
  \hline
   SXP ID & Telescope & Date & Exposure Time (s) & H$\alpha$ EW (\AA) & H$\beta$ EW (\AA)\\
 \hline
 SXP101 & SALT Red & - & - & -& -\\
        & SALT Blue & 10 Oct 2006 & 450 & -& $0.0\pm0.7$\\
        & ESO & 14 Sep 2006 & 2400 & -& $1.1\pm0.2$\\
        & AAT & 31 Aug 2006 & 5400 & $0.2\pm0.3$& $4.7\pm0.3$\\
        & SAAO & 07 Sep 2004 & 2$\times$1000 & $1.7\pm0.2$& -\\
        & SAAO & 10 Nov 2006 & 2$\times$1500 & $-2.9\pm0.5$& -\\
SXP700 & SALT Red & 15 Oct 2006 & 600 & $-13.4\pm0.1$& -\\
        & SALT Blue & 16 Oct 2006 & 600 & -& $-1.0\pm0.1$\\
        & ESO & 13 Sep 2006 & 2400 & -& $-0.7\pm0.1$\\
        & AAT & 31 Aug 2006 & 5400 & $-13.4\pm0.2$& $-0.8\pm0.2$\\
        & SAAO & 10 Nov 2006 & 1800 & $-12.9\pm0.4$& -\\
SXP348 & SALT Red & - & - & -& -\\
        & SALT Blue & - & - & -& -\\
        & ESO & 14 Sep 2006 & 1800 & -& $0.3\pm0.2$\\
        & AAT & - & - & -& -\\
        & SAAO & 12 Sep 2004 & 2500 & $-8.7\pm0.9$& -\\
        & SAAO & 30 Oct 2005 & 1000 & $-2.9\pm0.6$& -\\
        & SAAO & 08 Nov 2006 & 1200 & $-6.3\pm0.5$& -\\
SXP65.8 & SALT Red & - & - & -& -\\
        & SALT Blue & - & - & -& -\\
        & ESO & 13 Sep 2006 & 2000 & -& $1.2\pm0.1$\\
        & AAT & 31 Aug 2006 & 2$\times$5400 & $-21.3\pm0.3$& $0.9\pm0.2$\\
        & SAAO & 13 Nov 2006 & 1800 & $-13.5\pm0.6$& -\\
 \hline
\end{tabular}\\
\end{minipage}
\end{table*}

\section{Observations}
\subsection{Classification Method}
Spectral classification of Be stars in the SMC is particularly difficult.  Classification of Be stars in the Galaxy relies on using the ratios of many metal lines \citep{wal90}, unfortunately the metallicity of SMC stars is lower than those in the Galaxy and so these metal lines appear extremely weak and are often not present.  There is also the added difficulty that the Balmer lines in particular will be rotationally broadened and hence obscure any comparisons to neighbouring lines. Other difficulties are added when we consider the effect of the circumstellar disk.  We often get H$\alpha$ emission originating from the circumstellar disk but we can also see the circumstellar disks effect on the higher order balmer lines though in-filling.  As a result we use the classification criteria and methods as set out by \citet{len97} and \citet{eva04}. 

For the luminosity classification we have adopted the classification method set out in \citet{wal90}, however since this method relies on line ratios we face all the same difficulties as previously mentioned.  We have also performed a check on this luminosity classification by comparing the absolute magnitude of the source in the V-band with the spectral classification obtained.  Here we have adopted a distance modulus for the SMC of 18.9 \citep{{har03}} and we use the relevant tables in \citet{weg06}. These tables are based on absolute magnitudes from HIPPARCOS data.  Although the luminosities of Be stars in the SMC may differ somewhat to those in the Milky Way we have adopted this method as a check and recognise that in some cases the results obtained may be uncertain.

\begin{figure*}
 \includegraphics[width=120mm, angle=90]{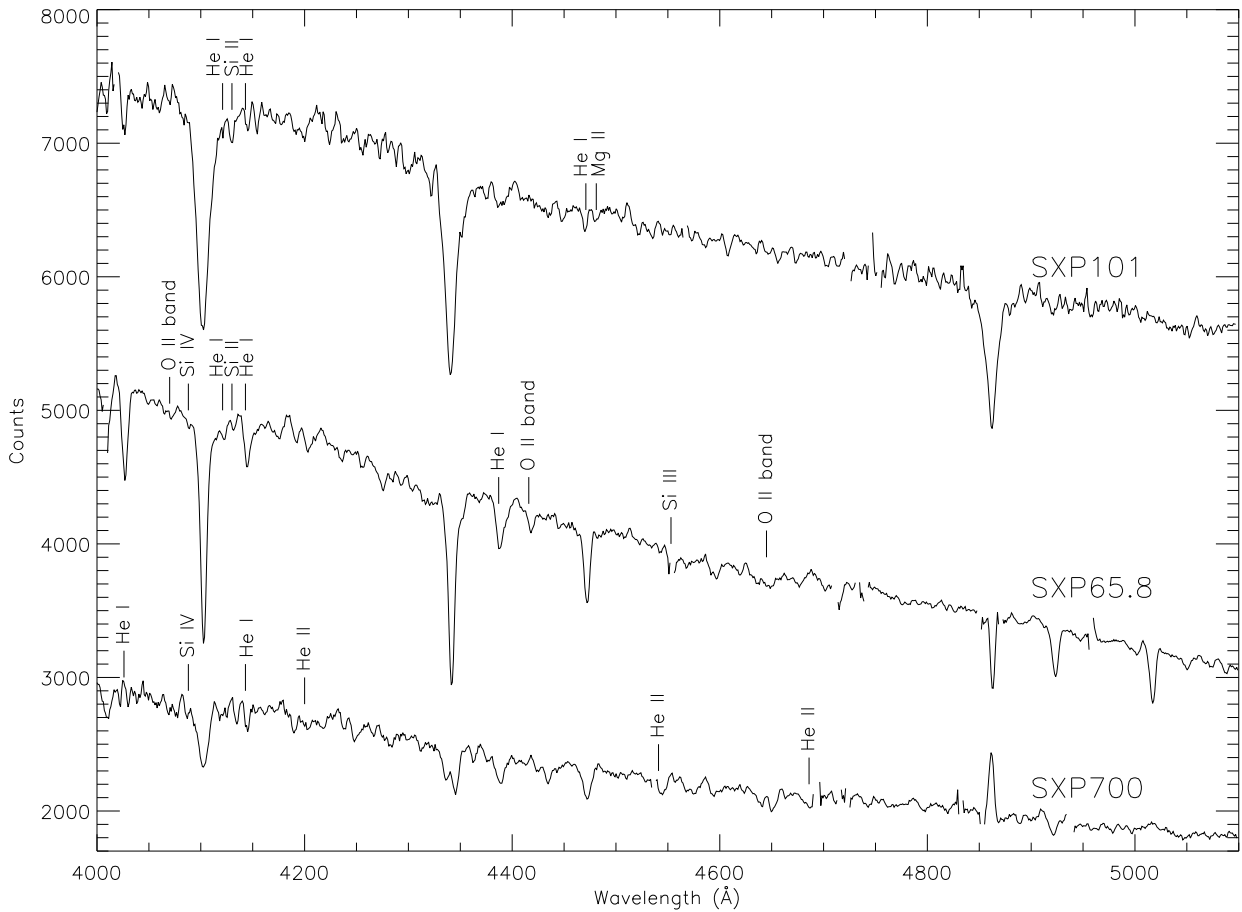}
  \caption{AAT spectra taken on 31 Aug 2006. Top to bottom: SXP101, SXP65.8, SXP700, (smoothed with gaps for dead pixels).\label{fiaat}}
\end{figure*}
\begin{figure*}
 \includegraphics[width=120mm, angle=90]{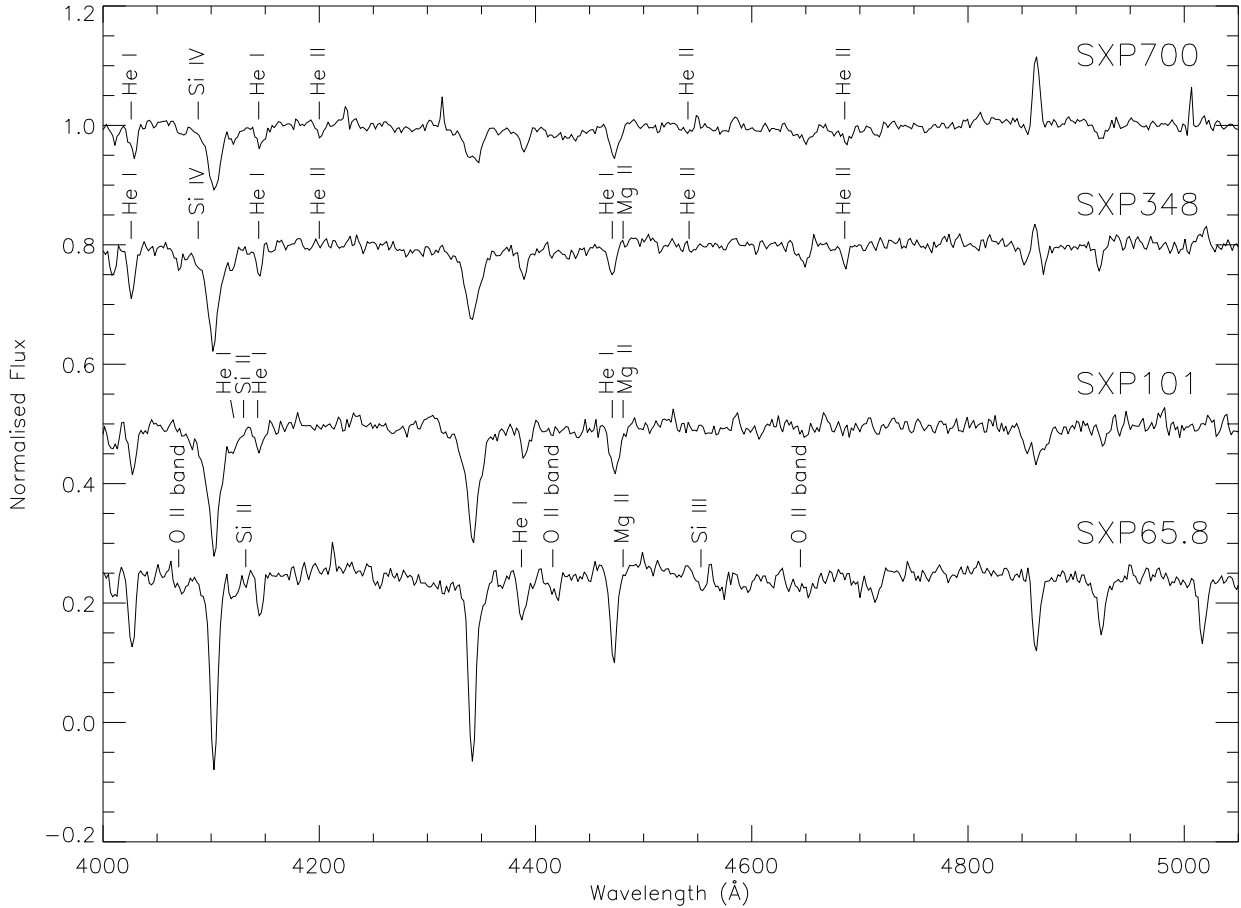}
  \caption{ESO blue spectra taken on 14 Sep 2006. Top to bottom: SXP700, SXP348, SXP101, SXP65.8 (no smoothing).\label{fieso}}
\end{figure*}
\begin{figure*}
 \includegraphics[width=120mm, angle=90]{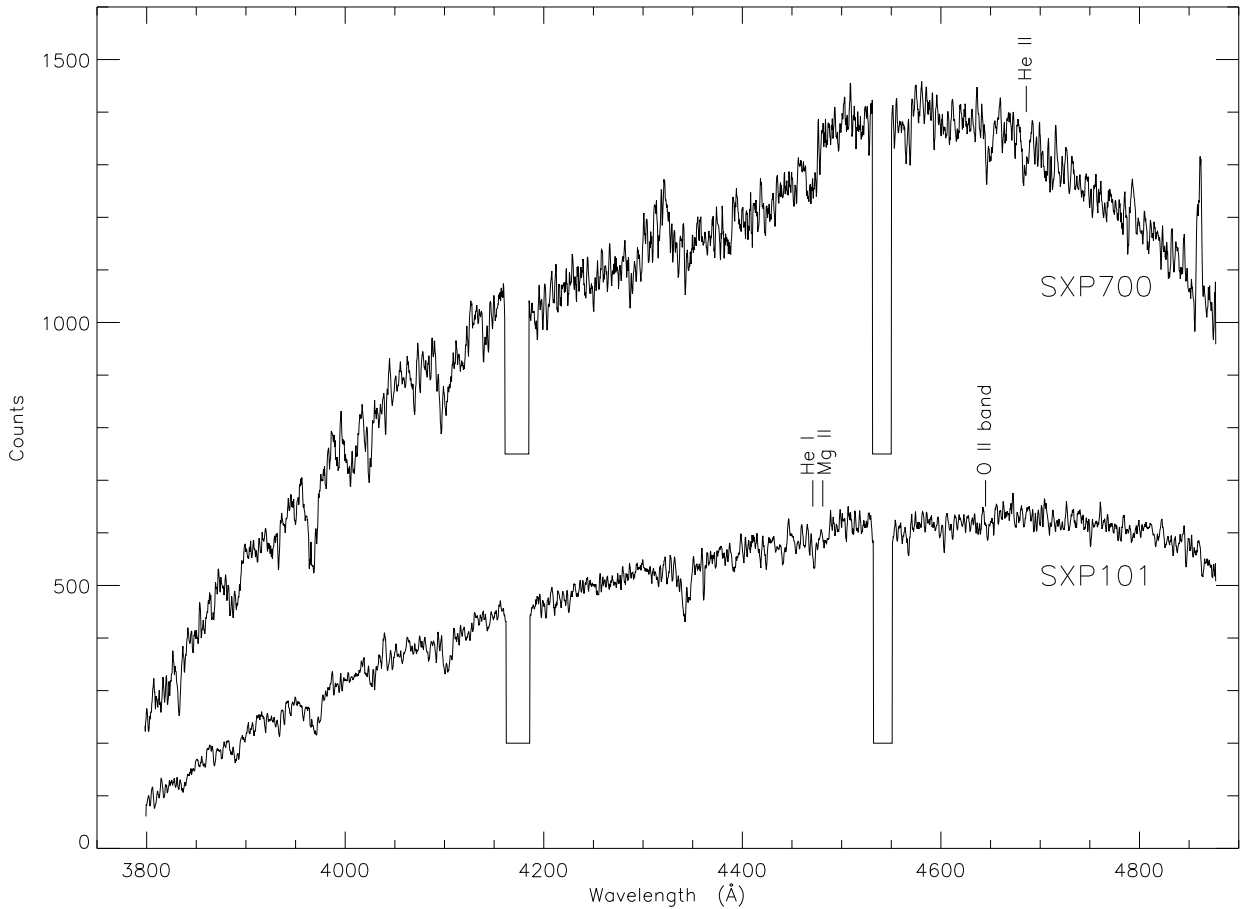}
  \caption{SALT blue spectra of SXP700 (16 Oct 2006) and SXP101 (10 Oct 2006) (chip gaps are clearly visible).\label{fisaltb}}
\end{figure*}

\subsection{SXP101}
This source was detected by both ROSAT (RX J0057.3-7325) and {\it ASCA} \citep[AX J0057.4-7325;][]{kah99,tor00}.  Coherent pulsations were first detected in the {\it ASCA} data at a period of $101.45\pm0.07$\,s.  The resultant overlapping error circles allowed for some counterparts to be tentatively assigned \citep{edg03}.  The detection of pulsations at $101.16\pm0.26$\,s in the {\it Chandra} data \citep{mcg07} has allowed the counterpart to be clearly identified as the source previously labelled E in Figure~3 of \citet{edg03}, MACS J0057-734 10 \citep{tuc96}.  This source had first been identified as a possible counterpart in January 2000 due to its r-H$\alpha$ colour revealing an excess of H$\alpha$.  Subsequently this source was observed 4 years later in 2004 using the 1.9m SAAO telescope, two spectra taken on the night have been co-added and are shown in Figure~\ref{fi101ha}.  We can clearly see the H$\alpha$ line to be in absorption, hence the source was not considered to be a prime candidate.  Strangely the AAT spectrum (Figure~\ref{fi101ha}) shows neither absorption or emission in H$\alpha$.  In fact it is perfectly filled in. Curiously the higher order Balmer lines in the AAT spectrum (Figure~\ref{fiaat}), show no signs of infilling, they are all clearly in absorption with no peculiar features.  A third red spectrum (Figure~\ref{fi101ha}) was taken recently at SAAO in November 2006. Two spectra were taken consecutively and have been co-added to increase the signal to noise.  The H$\alpha$ profile now appears to be slightly in emission, suggesting its Be nature. There have been several observations of the H$\beta$ emission line during the last few months of 2006 (Figure~\ref{fiaat},~\ref{fieso},~\ref{fisaltb}).  These observations appear to show that in the course of little more than one month the line has completely filled itself in.  

Classification has been made using blue spectra taken by AAT, ESO and SALT (Figures~\ref{fiaat},~\ref{fieso},~\ref{fisaltb}).  The counterpart is extremely difficult to classify due to a severe lack of metallic lines.  No HeII lines are visible hence the counterpart must be B1 or later \citep{len97}.  There is some evidence for the presence of the OII 4640-50\AA\ band in the SALT spectra however it seems to be lacking in the other spectra.  Comparison of all three spectra with those given in Figure~4 of \citet{eva04} would seem to place the counterpart in a broad spectral range of B1-B5. MgII 4481\AA\ is possibly present in all three spectra in particular the SALT and AAT ones, however the resolution of the ESO spectra is not sufficient to deblend the line from the neighbouring HeI line.  If the presence of this line is to be believed, then combined with the lack of SiIII 4553\AA\ allows a lower classification limit of B3 to be made \citep{len97}.  Visable in the AAT spectra is the SiII blend of 4128/4132\AA\ and HeI line at 4143\AA.  HeI 4121\AA\ is possibly visible on the red shoulder of the H$\delta$ line. However, it is heavily obscured and cannot be used for classification. The ratio of SiII to HeI line would suggest a classification around B8 \citep{eva04}. This line ratio is not seen in the ESO spectra, in fact SiII is barely seen and the HeI line appears strongly suggesting a classification more like B5. A classification range of B3-B5 would agree with the Balmer lines appearing very broad and deep which is typical of a mid range Be star.   The luminosity classification method of \citet{wal90} only goes as far as classifying B2 stars, hence we will only use the estimate from the V-band magnitude.  We would estimate the luminosity class to be about Ib-II from Table~10 presented in \citet{weg06}.  A luminosity class of Ib-II would make the counterpart a supergiant and not a Be star.  This luminosity classification should be treated cautiously due to the methods used. 

\subsection{SXP700}
\begin{figure}
 \includegraphics[width=60mm, angle=90]{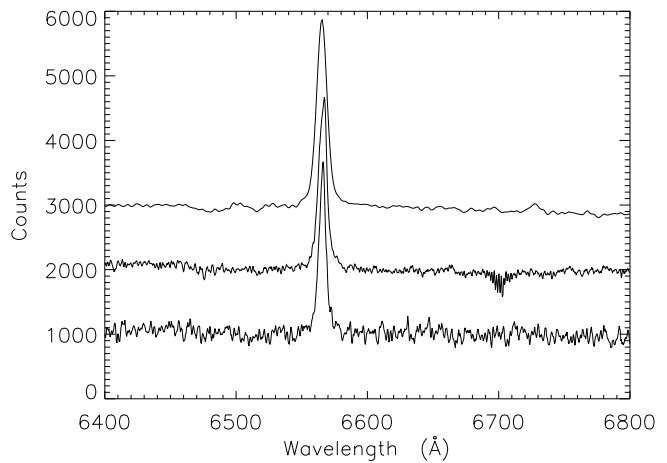}
  \caption{H$\alpha$ spectra of SXP700. Spectra dates top to bottom: 31 Aug 2006, 15 Oct 2006, 10 Nov 2006 (this spectrum has been scaled to allow the comparison of the H$\alpha$ profile).\label{fi700ha}}
\end{figure}

SXP700 is one of the new pulsars detected on 06 February 2006 in the SMC wing survey.  The {\it Chandra} source is coincident with [MA93] 1301 \citep{mey93}.  Three red follow up spectra taken within a few months of each other by SALT, AAT and SAAO reveal the counterpart star to have a very strong H$\alpha$ line in emission (Figure~\ref{fi700ha}).  This strongly suggests the identification of this object as a Be/X-ray binary.  The three observations were roughly spaced a month apart from each other. It can be seen in Table~\ref{taobs} that the equivalent width of H$\alpha$ has barely changed over the course of these observations indicating that the circumstellar disk was in a fairly stable state.

Classification of the counterpart has been possible through the three blue spectra taken by AAT, ESO and SALT (Figures~\ref{fiaat},~\ref{fieso},~\ref{fisaltb}).  Weak HeII at 4686\AA\ can be seen in all the spectra; using the classification guide set out by \citet{eva04} we can immediately place this object as earlier than B1.  No other HeII lines are visable in either the AAT or SALT spectra however, HeII 4200\AA\ is present in the ESO spectra.  This combined with the lack of HeII 4541\AA\ allows the classification of B0-B0.5 to be made \citep{eva04}. H$\beta$ can also be seen to be strongly in emission in all three blue spectra.  A luminosity classification can be made by comparing the ratio of SiIV 4089\AA\ to both HeI 4026\AA\ and 4144\AA.  The ESO spectra shows strong HeI lines compared with slight evidence of SiIV in the blue shoulder of the H$\delta$ line.  Using the ratio of these lines results in a luminosity class of V \citep{wal90}.  Using the method based upon the V-band magnitude yields a classification of III-V.

\subsection{SXP348}
\begin{figure}
 \includegraphics[width=60mm, angle=90]{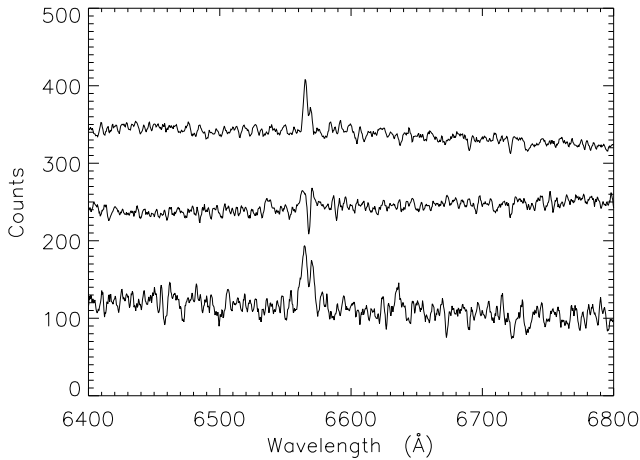}
  \caption{H$\alpha$ spectra of SXP348. Spectra dates top to bottom: 12 Sep 2004, 30 Oct 2005, 08 Nov 2006.\label{fi348ha}}
\end{figure}

This is the only detected pulsar from the {\it Chandra} Wing survey with a previously known counterpart. It was first detected by {\it BeppoSAX} in 1998 \citep{isr98} with a pulse period of $345.2\pm0.1$\,s. The {\it Chandra} source is coincident with [MA93] 1367 \citep{mey93}, a Be star \citep{hug94,isr98}.  Spectra taken in 2004 and 2005 by the 1.9m SAAO telescope (Figure~\ref{fi348ha}) show that the H$\alpha$ line has been varying over recent years.  In 2004 the line was strongly in emission with a small second peak on its red shoulder, just over a year later in 2005 the line appears to have dropped in strength and displays a shell profile. A recent spectrum (Figure~\ref{fi348ha}) taken on 08 November 2006 shows H$\alpha$ to be back in emission, this time with a slightly more prominent feature on the red shoulder making it appear more double peaked.  The ESO spectrum (Figure~\ref{fieso}) taken near to this most recent red spectrum shows the H$\beta$ line to be largely filled in with an emission peak at its centre.  The H$\alpha$ profile has also shrunk in size quite dramatically since it was observed in 1992 \citep{hug94}, they report an equivalent width of -22\AA\ however they did not note anything about the physical shape of the emission.  The H$\beta$ line has also reduced in size since 1992, where they reported an equivalent width of -1.7\AA.

From the ESO spectra (Figure~\ref{fieso}) we can see a strong HeII line at 4686\AA; this firmly restricts the spectral class to earlier than B1.  Further constraints can be made based on the presence of very weak HeII at 4541\AA\ and 4200\AA\ in the ESO spectra. A spectral classification of B0-B0.5 can be placed on this counterpart \citep{eva04}. This agrees with the spectral range of O9-B1~(V-III) determined in \citet{hug94}, where they estimate the spectral class to around B0 based on the strength of HeI 4471\AA\ relative to its neighbour MgII 4481\AA. The presence of HeII 4686\AA\ was uncertain in their spectrum due to a flat fielding correction.  Using the same HeI and SiIV lines in the ESO spectrum for the luminosity classification as used for SXP700 yields a very similar situation, strong HeI lines and a moderate bump representing SiIV on the shoulder of H$\delta$.  Using this ratio a luminosity classification of V is made \citep{wal90}, this is consistent with the range of III-V estimated from the V-band magnitude.

\begin{figure}
 \includegraphics[width=60mm, angle=90]{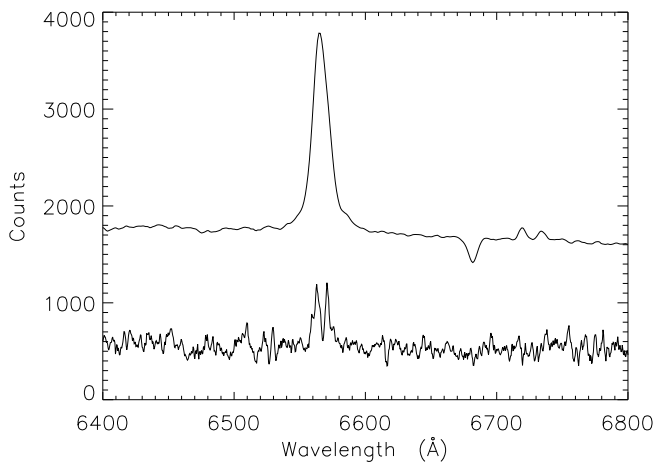}
  \caption{H$\alpha$ spectra of SXP65.8. Spectra dates top to bottom: 31 Aug 2006, 13 Nov 2006 (this spectrum was scaled to allow the comparison of the H$\alpha$ profile).\label{fi65ha}}
\end{figure}

\subsection{SXP65.8}
\subsubsection{Optical}
SXP65.8 is the second of the new pulsars detected in the SMC wing survey on 10 February 2006. The {\it Chandra} source is coincident with the emission line star [MA93] 1619 \citep{mey93}.  Follow-up red spectra were taken by SAAO and AAT (Figure~\ref{fi65ha}).  The counterpart was observed twice with the AAT on the same night, roughly one hour apart.  Each observation was 5400s long.  Since these spectra are both of high quality and show no astrophysical difference we have only included one of them here. The strong H$\alpha$ emission line present in the AAT spectra strongly suggests that the counterpart is indeed a Be/X-ray binary. In the few months between the AAT and SAAO spectra the H$\alpha$ line has significantly modified its profile to a double peaked structure.  This would indicate that the circumstellar disk has shrunk in size and now the absorption in the stellar photosphere is becoming a more dominant effect.

The classification for this counterpart has been based on AAT and ESO blue spectra (Figures~\ref{fiaat},~\ref{fieso}).  Clear OII absorption bands visible at 4640-50\AA, 4415-17\AA\ and 4070\AA are present in both spectra, the HeII lines have also completely disappeared.  These allow a classification of B1-B3 to be made \citep{eva04}.  Several silicon lines can help to refine the classification.  The presence of a strong SiIII 4553\AA\ line in the AAT spectrum is dubious since it falls on a dead pixel, however, the continuum does appear to take a strong dip just before and then rise just after as if the line were present.  Also visible in the AAT spectrum is weak SiIV 4088\AA and SiII 4132\AA.  The ESO spectra confirms the presence of SiIII and SiII, however SiIV only appears as a slight bump on the shoulder of H$\delta$.  There is the possibility of very weak MgII 4481\AA\ absorption seen on the shoulder of the HeI line in the ESO spectra.  Comparing the strengths of SiIII to MgII we can place the object as earlier than B2 \citep{len97}.  Using the detections of the other lines we can also say that OII is stronger than SiIV 4088\AA\ and hence refine the classification to B1-B1.5. SiIV 4116\AA\ is unresolvable in both spectra due to the width of the H$\delta$ feature.  We can not be certain of its presence or absence, hence a final classification of B1-B1.5 is made.  There are also notable FeII lines in absorption at 4924\AA\ and 5018\AA. Using the ratio of SiIII to HeI 4387\AA\ we can make a luminosity classification of II-III \citep{wal90} this agrees with that obtained using the magnitude method.

\subsubsection{X-ray}
\begin{figure}
 \includegraphics[width=60mm, angle=90]{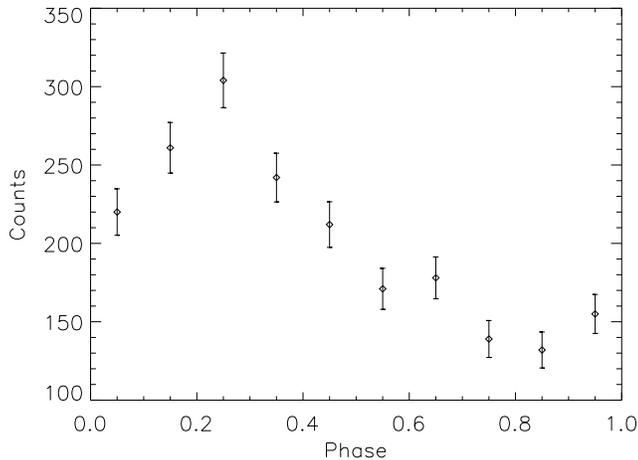}
  \caption{Pulse profile for SXP65.8 in the energy range 0.5-8.0\,keV.\label{filc}}
\end{figure}

Using {\it Chandra} data \citep[for more details see][]{mcg07} we have extracted phase resolved spectra for SXP65.8 using CIAO~v3.4 standard tools.  The spectra were regrouped by requiring at least 10 counts per spectral bin.  The subsequent spectral fitting and analysis were performed using XSPEC~v12.3.0.  A phase binned light curve was created (Figure~\ref{filc}) based on MJD\,53776.82172 and a pulse period of 65.78s \citep{mcg07}. The two phase binned spectra were extracted using the phase ranges 0.0-0.5 and 0.5-1.0.  We fitted each of these with an absorbed power law, and fixed the column density at the value for the SMC of $6\times10^{20}$\,cm$^{-2}$ \citep{dic90}. The low phase spectrum is well fit with a $\Gamma=0.34\pm0.08$ with a reduced $\chi^2=0.85$, and the high phase spectrum is well fit with $\Gamma=0.52\pm0.10$ with a reduced $\chi^2=1.42$.  These values are consistent right at the extremes of their errors. However, they would also seem to indicate that the pulsed emission from the beam is harder than the persistent emission.  The pulse fraction has been calculated for both soft and hard energy bands (0.5-2.0\,keV and 2.0-8.0\,keV respectively). The values were calculated by taking $(F_{max}-F_{min})/(F_{max}+F_{min})$ where $F_{max}$ and $F_{min}$ are the maximum and minimum points of the folded light curve for each energy band. $PF_{S}=34\pm9$\,\% and $PF_{H}=43\pm6$\,\%. The values are both conisitant within errors and with the value in \citet{mcg07}. This is probably due to the lack of soft counts.

\section{Discussion}
\subsection{SXP101}
The identification of the correct counterpart to SXP101 has finally enabled its identification as a Be/X-ray binary.  The H$\alpha$ and H$\beta$ line profiles provide strong evidence to suggest that the circumstellar disk is an extremely variable component of the system.  The most recent observations made at the end of 2006 suggest that the circumstellar disk has recently grown in size. Interestingly the H$\beta$ line has shown some considerable variation over the course of a few months.  Figure~\ref{fiaat} clearly shows all the Balmer lines to be deep and well defined including H$\beta$; 14 days later and the H$\beta$ line has considerably changed and is almost totally filled in with the remaining Balmer lines showing no change (Figure~\ref{fieso}). This would agree with the H$\alpha$ observations which show that over approximately the same time scale the H$\alpha$ line has gone from being filled in completely to slightly in emission. These recent observations provide strong evidence to suggest that the circumstellar disk is an extremely variable object.  The apparent lag of the H$\beta$ line with respect to H$\alpha$ could be evidence of the emission originating from a different part of the circumstellar disk than the H$\alpha$.  This could be due to temperature or density changes.  Further modelling of the circumstellar disk structure would be needed to confirm this.

The classification of B3-B5 is also potentially very interesting since it would make this Be/X-ray binary one of the latest types to have been found.  Comparing the spectral type of SXP101 with the spectral distribution found in the Galaxy \citep{neg98}, clearly shows that SXP101 would lie on its own, becoming the latest spectral type found. McBride et al. (2007, in preparation) will present full spectral classifications and analysis of all the Be/X-ray binaries in the SMC where the counterparts are known.  This work will help in understanding any potential differences between the spectral distribution in the SMC and the Galaxy.  

\subsection{SXP700}
The presence of a strong H$\alpha$ emission line and a classification of B0-B0.5 clearly places this Be/X-ray binary right in the centre of the Galactic spectral distribution \citep{neg98}.  The unchanging equivalent widths of the H$\alpha$ line (Table~\ref{taobs}) suggests that the circumstellar disk was in a fairly stable state.  An orbital period of $267.38\pm15.10$\,d was found in OGLE III data \citep{mcg07}.  This period is entirely consistent with that predicted from the Corbet diagram \citep{cor86} for a 700\,s pulsar.  Assuming SXP700 only goes into outburst everytime the neutron star passes through periastron then we would only expect outbursts approximately every 9 months.  For such long intervals where no accretion is taking place the circumstellar disk would have ample time to grow to a stable state.  Our observations taken during the orbital phase range 0.41-0.68 \citep[see Figure 6.][]{mcg07}, show that over approximately one quarter of this orbital period the circumstellar disk has remained stable.  This could be indicative of a stable truncated circumstellar disk at its maximum equivalent width, similar to the stable circumstellar disk observed in GRO\,J1008-57 \citep[][]{coe07}.  \citet{rei07} presents an up to date version of the P$_{orb}$-EW(H$\alpha$) diagram containing all the known Galactic and Magellanic Be/X-ray binaries where the orbital periods are known. The measured H$\alpha$ value is assumed to be representative of a circumstellar disk at maximum size.  The measured values for SXP700 are not consistent with this plot.  However, we also note that the values for GRO\,J1008-57 \citep[][]{coe07} where the circumstellar disk is in a stable state are also not consistent with the plot.  In order to properly place SXP700 on this diagram we would need many more observations at varying times to evaluate the changes that take place in the circumstellar disk.

\subsection{SXP348}
\begin{figure}
 \includegraphics[width=60mm, angle=90]{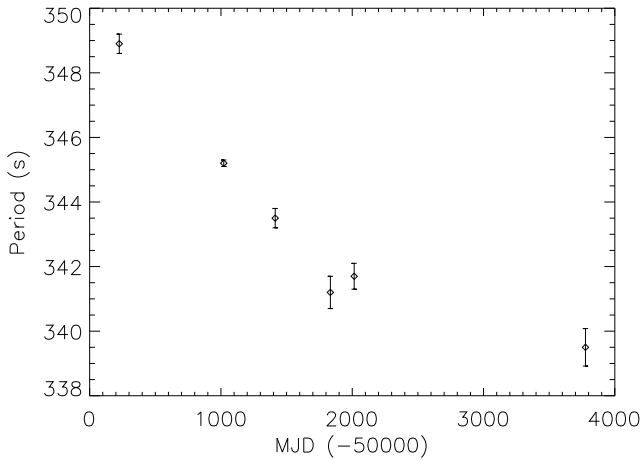}
  \caption{Spin period of SXP348 over the past 10 years.\label{fi348sp}}
\end{figure}

This source was first detected by {\it BeppoSAX} in 1998 \citep{isr98} with a pulse period of $345.2\pm0.1$\,s.  After its discovery a previous {\it ASCA} observation from 1996 was found to have the same source present at a slightly slower spin period of $348.9\pm0.3$\,s \citep{yok98}. Several later observations reveal that SXP348 has continued its spin up at an almost constant rate, detections by {\it Chandra} in 1999 of $343.5\pm0.3$\,s pulsations \citep{isr00} and then {\it XMM-Newton} observations in 2000 at $341.21\pm0.5$\,s \citep{hab04}. An {\it XMM-Newton} observation in 2001 showed a slight spin down of the pulsar to $341.7\pm0.4$\,s \citep{sas03}.  The recent {\it Chandra} observation at $339.56\pm0.58$\,s supports this trend of continued spin-up.  Figure~\ref{fi348sp} shows how SXP348 has been spinning up \citep{mcg07} over the last 10 years.  The recent {\it Chandra} observation suggests that the constant spin up rate has recently slowed.

It appears that SXP348 has transitioned at some point into a different state where the spin period is now changing much more slowly.  We have divided the observations into two epochs, with the transition point being the average of the two {\it XMM-Newton} observations that fall around MJD\,52000.  Epoch one yields $\dot{P}_{1}=(5.1\pm0.4)\times10^{-8}$\,ss$^{-1}$, which is consistant with the value found in \citet{hab04}, this implies $L_{x1}=(4.5\pm0.4)\times10^{35}$\,erg\,s$^{-1}$.  Epoch two yields $\dot{P}_{2}=(1.2\pm0.5)\times10^{-8}$\,ss$^{-1}$, which implies $L_{x2}=(1.1\pm0.4)\times10^{35}$\,erg\,s$^{-1}$.  We have derived an estimate on the magnetic field strength using Eq.~(6.24) of \citet{fra02}.  The luminosity value of epoch one yields a field strength of $B_{1}=(4.3\pm1.6)\times10^{13}$\,G.  The field strength derived from epoch two is consistent with this value but has significantly larger errors.  The change in luminosity is as expected and would indicate that SXP348 has had significant changes to its accretion rate, possibly implying that the pulsar is nearing a spin period where the accretion flow will be cut off due to the propeller effect caused by the magnetosphere.  The estimated value of the magnetic field is higher than expected, this is probably a reflection of the estimation method used when only a handful of points are known.

The classification of B0-B0.5 agrees with the spectral type of O9-B1\,(V-III) determined by \citet{hug94}, and places this source in the middle of the Galactic spectral distribution.  The variation of the H$\alpha$ profile shows that the circumstellar disk is variable on time scales of around a year, this is twice the expected orbital period as predicted from the Corbet diagram \citep{cor86}.  Since the state of the circumstellar disk is intricately linked to the spin period changes we would require many more observations to fully understand the dynamics of the system. 

\subsection{SXP65.8}
Observations of a varying H$\alpha$ emission line and a classification of B1-B1.5 confirm SXP65.8 as a Be/X-ray binary. Since the H$\alpha$ emission is arising from the circumstellar disk we deduce that the reduction in equivalent width is due to the circumstellar disk reducing in size over the course of a few months. 

The phase resolved spectroscopy would appear to suggest that as the X-ray beam passes through the line of sight the X-ray emission becomes harder.  This would indicate that the soft emission is originating in a different location to the magnetically collimated beam and is probably more representative of thermal emission from either the surface of the neutron star or the accretion disk. 

\section{Conclusions}
Four pulsars were detected in the SMC {\it Chandra} wing survey, SXP101, SXP700, SXP348 and SXP65.8.  Follow up optical observations have enabled spectroscopic classifications of all the counterparts. All the counterparts are classified within the range B0-B5.  Red spectra covering the H$\alpha$ atomic line has revealed that all the sources have at some point in time exhibited H$\alpha$ emission, resulting in their classification as emission line stars.  Three of the Be/X-ray binaries fall within the known Galactic spectral distribution for Be/X-ray binaries \citep{neg98}.  SXP101 falls outside of this distribution and hence becomes the latest spectral type known for a Be/X-ray binary.  Currently there is not enough evidence to suggest that there is a different population in the Wing compared to the Bar.  The work by McBride et al. (2007, in preparation) will shed some light on the spectral distribution of Be/X-ray binaries in the SMC.  From the H$\alpha$ monitoring it would appear that SXP700 has a stable circumstellar disk, unlike the three other sources where variations in the profile would suggest a more dynamic circumstellar disk.  This is likely to be due to the differences in the length of the orbital periods and hence the influence of the neutron star on the circumstellar disks geometry.  

\citet{mcg07} suggest that the Wing pulsars could be coming from a different population to the Bar pulsars due to the pulsars exhibiting harder X-ray spectra, however, our findings are unable to clarify this matter any further.

\section*{Acknowledgements}
We wish to thank all the SALT astronomers in particular, Martin Still, Petri Vaisanen and Alexei Kniazev for their help in observing the targets and in answering many questions. We thank all the staff at the AAT in particular Rob Sharp for his help with the data processing.

\bsp

\label{lastpage}


\begin{thebibliography}{99}
\bibitem[\protect\citeauthoryear{Coe et al.}{2005}]{coe05}
Coe M.J., Edge W.R.T., Galache J.L., McBride V.A., 2005, MNRAS, 356, 502
\bibitem[\protect\citeauthoryear{Coe et al.}{2007}]{coe07}
Coe M.J., et al., 2007, MNRAS, 378, 1427
\bibitem[\protect\citeauthoryear{Corbet}{1986}]{cor86}
Corbet R.H.D., 1986, MNRAS, 220, 1047
\bibitem[\protect\citeauthoryear{Dickey \& Lockman}{1990}]{dic90}
Dickey J.M., Lockman F.J., 1990, A\&A, 28, 215
\bibitem[\protect\citeauthoryear{Edge \& Coe}{2003}]{edg03}
Edge W.R.T., Coe M.J., 2003, MNRAS, 338, 428
\bibitem[\protect\citeauthoryear{Evans et al.}{2004}]{eva04}
Evans C.J., Howarth I.D., Irwin M.J., Burnley A.W., Harries T.J., 2004, MNRAS, 353, 601
\bibitem[\protect\citeauthoryear{Frank et al.}{2002}]{fra02}
Frank J., King A., Raine D.J., 2002, Accretion Power in Astrophysics: Third Edition. Cambridge University Press
\bibitem[\protect\citeauthoryear{Haberl \& Pietsch}{2004}]{hab04}
Haberl F., Pietsch W., 2004, A\&A, 414, 667
\bibitem[\protect\citeauthoryear{Harries et al.}{2003}]{har03}
Harries T.J., Hilditch R.W., Howarth I.D., 2003, MNRAS, 339, 157
\bibitem[\protect\citeauthoryear{Hughes \& Smith}{1994}]{hug94}
Hughes J.P., Smith R.C., 1994, AJ, 107, 4
\bibitem[\protect\citeauthoryear{Israel et al.}{1998}]{isr98}
Israel G.L., Stella L., Campana S., Covino S., Ricci D., Oosterbroek T., 1998, IAUC 6999, 1
\bibitem[\protect\citeauthoryear{Israel et al.}{2000}]{isr00}
Israel G.L., et al., 2000, ApJ, 531, 131
\bibitem[\protect\citeauthoryear{Kahabka et al.}{1999}]{kah99}
Kahabka P., Pietsch W., Filipovic M.D., Haberl F., 1999, A\&AS, 136, 81
\bibitem[\protect\citeauthoryear{Lennon}{1997}]{len97}
Lennon D.J., 1997, A\&A, 317, 871
\bibitem[\protect\citeauthoryear{Massey}{2002}]{mas02}
Massey P., 2002, ApJS, 141, 81
\bibitem[\protect\citeauthoryear{McGowan et al.}{2007}]{mcg07}
McGowan K.E., et al., 2007, MNRAS, 376, 759
\bibitem[\protect\citeauthoryear{Meyssonnier \& Azzopardi}{1993}]{mey93}
Meyssonnier N., Azzopardi M., 1993, A\&AS, 102, 451
\bibitem[\protect\citeauthoryear{Negueruela}{1998}]{neg98}
Negueruela I., 1998, A\&A, 338, 505
\bibitem[\protect\citeauthoryear{Reig}{2007}]{rei07}
Reig P., 2007, MNRAS, 377, 867
\bibitem[\protect\citeauthoryear{Sasaki et al.}{2003}]{sas03}
Sasaki M., Pietsch W., Haberl F., 2003, A\&A, 403, 901
\bibitem[\protect\citeauthoryear{Torii et al.}{2000}]{tor00}
Torii K., Kohmura T., Yokogawa J., Koyama K., 2000, IAUC, 7441, 2
\bibitem[\protect\citeauthoryear{Tucholke et al.}{1996}]{tuc96}
Tucholke H.-J., De Boer K.S., Seitter W.C., 1996, A\&AS, 119, 91
\bibitem[\protect\citeauthoryear{Walborn \& Fitzpatrick}{1990}]{wal90}
Walborn N.R., Fitzpatrick E.L., 1990, PASP, 102, 379
\bibitem[\protect\citeauthoryear{Wegner}{2006}]{weg06}
Wegner W., 2006, MNRAS, 371, 185
\bibitem[\protect\citeauthoryear{Yokogawa \& Koyama}{1998}]{yok98}
Yokogawa J., Koyama K., 1998, IAUC 7009, 3

\end{thebibliography}
\end{document}